\documentclass[sigconf,screen, natbib=false, 9pt]{acmart}
\newcommand{\floorset}{$\mathtt{FloorSet }$}
\AtBeginDocument{%
  }

\usepackage{microtype}
\usepackage{amsmath}
\usepackage{amsfonts}
\usepackage{placeins}
\usepackage{graphicx}
\usepackage{subcaption}
\usepackage{booktabs} 
\usepackage{algorithm}
\usepackage[noend]{algorithmic}
\usepackage{adjustbox}
\usepackage{xspace}
\usepackage{multirow}
\usepackage{epsfig}
\usepackage{graphics}
\usepackage{times}
\usepackage{algorithm}
\usepackage[noend]{algorithmic}
\usepackage{bm}

\usepackage{color}
\usepackage{array}
\usepackage{amsmath}
\usepackage{amsthm}
\usepackage{bm}
\usepackage{pifont}
\usepackage{comment}
\usepackage{psfrag}
\usepackage{verbatim,url}
\usepackage{float,tabularx}
\usepackage{tabularx}
\usepackage{float}

\usepackage{caption}
\usepackage{subcaption}
\usepackage{setspace}

\usepackage{pifont}
\usepackage{hyperref}

\RequirePackage[
  datamodel=acmdatamodel,
  style=acmnumeric,
  ]{biblatex}

\begin{document}


\title{FloorSet - a VLSI \underline{Floor}planning Data\underline{set} with Design Constraints of Real-World SoCs} 

\author{Uday Mallappa}
\affiliation{%
  \institution{Intel Labs} 
  \country{USA}  
}

\author{Hesham Mostafa}
\affiliation{%
  \institution{Intel Labs} 
  \country{USA}  
}

\author{Mikhail Galkin}
\affiliation{%
  \institution{Intel Labs} 
  \country{USA}  
}

\author{Mariano Phielipp}
\affiliation{%
  \institution{Intel Labs} 
  \country{USA}  
}

\author{Somdeb Majumdar}
\affiliation{%
  \institution{Intel Labs} 
  \country{USA}  
}


\begin{abstract}
Floorplanning for systems-on-a-chip (SoCs) and its sub-systems is a crucial and non-trivial step of the physical design flow. It represents a difficult combinatorial optimization problem. A typical large scale SoC with 120 partitions generates a search-space of $\sim10^{250}$.  As novel machine learning (ML) approaches emerge to tackle such problems, there is a growing need for a modern benchmark that comprises a large training dataset and performance metrics that better reflect real-world constraints and objectives compared to existing benchmarks. To address this need, we present \floorset \: - two comprehensive datasets of synthetic fixed-outline floorplan layouts that reflect the distribution of real SoCs. Each dataset has $1M$ training samples and $100$ test samples where each sample is a synthetic floorplan.
\floorset-Prime comprises fully-abutted rectilinear partitions and near-optimal wire-length. 
A simplified dataset that reflects early design phases, \floorset-Lite  comprises rectangular partitions, with $<5\%$ white-space and near-optimal wire-length. Both datasets define hard constraints seen in modern design flows such as shape constraints, edge-affinity, grouping constraints, and pre-placement constraints. \floorset \ is intended to spur fundamental research on large-scale constrained optimization problems. Crucially, \floorset \ 
alleviates the core issue of reproducibility in modern ML driven solutions to such problems. \floorset \; is available as an open-source repository for the research community\footnote{\url{https://github.com/IntelLabs/FloorSet}}. 

\end{abstract}





\maketitle
\section{Introduction}
\label{sec:intro}
\begin{figure}[!h]
    \centering
\includegraphics[scale=0.4]{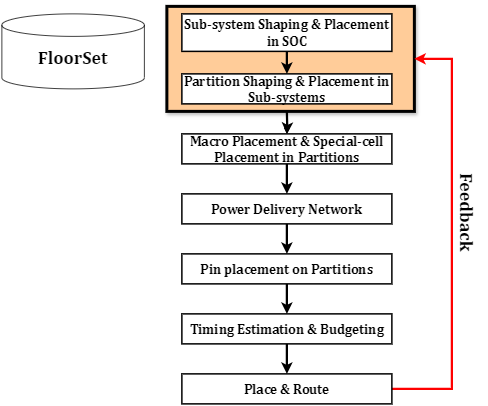}
\caption{Our work focuses on establishing realistic benchmarks \floorset, for the first two steps (shaded) of the design planning phase of the back-end flow.}
    \label{fig:floorset_pd}
\end{figure}

Circuit partitioning is the first step of the back-end physical design flow. It divides a flat, and large circuit netlist into more manageable partitions. This step defines area budgets specific to each partition, inter-partition connectivity constraints, connections to external terminals, and the respective positions of these external terminals. These outcomes define the requirements and constraints of the floorplanning step. Furthermore, the floorplanning task is governed by numerous placement constraints on a subset or all of the partitions. 
The goal of the floorplanning step is to determine optimal physical positions and shapes of the individual partitions comprising the SoC or a subsystem. The output from the floorplanning step serves as the starting point for the remainder of the physical design flow. Figure \ref{fig:floorset_pd} shows the typical back-end design steps - our work directly addresses the first two steps of partition and sub-system placement. 

\noindent \textbf{Floorplan Constraints:} The key challenge of the floorplanning formulation lies in the requirement to satisfy hard design constraints. Some important constraints are:
\begin{itemize}
    \item    Outline-constraint: Modern ASIC design relies on hierarchical (top-down) floorplanning, with a "fixed-outline" constraint for the SoC and its sub-systems. A floorplan optimized for area without accounting for this constraint may fail to fit within the specified outline.
    \item Shape-constraints: These specify the acceptable range of width-to-height ratios of each partition's bounding box. 
    \item Boundary-constraints: These specify that a partition must align with a specific edge or corner of the floorplanning outline. This alignment is driven by external interfacing requirements or system-level thermal considerations.
    \item Grouping-constraints: These specify a set of partitions that must be physically abutted - e.g.,  those operating on the same voltage or requiring simultaneous power-off.
     \item Multi-instantiation constraints: These define multiple partitions as instances of a shared main partition - e.g., eight instances of a CPU core partition. Thus, all such instances must share the same shape.
     \item Pre-placement constraints: These specify pre-defined locations of partitions that are often derived from past designs. 
\end{itemize}
\noindent \textbf{Supervised Learning and Data Paucity:}
Lately, there has been remarkable advancement in deep learning, that construct large parametric approximators by combining simpler functions. Deep learning demonstrates exceptional performance - particularly in scenarios involving high-dimensional spaces and abundant data points. Therefore, learning-based approaches can serve as promising alternatives for complex combinatorial optimization problems. However, the lack of sufficient training layouts is a major impediment in utilizing supervised learning methods for floorplanning. Due to this, prior learning-based techniques are often restricted to reinforcement learning (RL) \cite{kdd22,learnfp20, gplanner22, gfp22} to yield optimal floorplan layouts. However, such approaches struggle to simultaneously optimize the objectives and respect the hard design constraints for large-scale combinatorial optimization problems.  
This data paucity stems primarily from intellectual-property (IP) restrictions of the chip-design industry. Our work addresses this gap by generating a large corpus of synthetic layouts that are reflect the optimal placement statistics and design constraints of real, commercial SoCs. This dataset, \floorset, can serve as a comprehensive training and test set for ML based floorplanning algorithms.

\noindent \textbf{Overview:} The rest of the paper is organized as follows. \hyperref[sec:related]{Section 2} surveys the state of benchmarks and synthetic data in EDA. \hyperref[sec:methodology]{Section 3} describes our data generation pipeline, and the two dataset variants \floorset-Prime and \floorset-Lite. \hyperref[sec:experiments]{Section 4} summarizes the attributes of \floorset \: and presents the complexity of the generated data. Finally, conclusions are discussed in \hyperref[sec:conclusions]{Section 5}.

\section{Related Work}
\label{sec:related}
In recent years, an increasing number of works have applied ML to complex EDA problems 
\cite{mleda23, mlcad22, dpan21}. 
These works have targeted geometric problems \cite{binpacking21}, graph processing and optimization techniques\cite{timinggnn22, dagsizer23}, automated decision making \cite{chipformer23, agnesia20, rlsizer21}, vision-based approaches\cite{pdn20, drc20} and natural language models \cite{chipnemo24, chipchat23}. 
Such ML-based solutions in the EDA flow mitigate the need for exhaustive design optimization iterations by judiciously pruning the solution space associated with sub-optimal design quality. The resulting advancements present significant strides in various design stages such as high-level logic design \cite{hls20, hls22}, circuit design \cite{ls19, lo18}, physical design \cite{gfp22, gancts19, gr19, maskplace22}, verification \cite{verif23, fault21}, and manufacturing aspects \cite{asml} of conventional chip-design methodologies, offering notable opportunities for improvement.  

\noindent \textbf{EDA Benchmarks:} 
There are several examples of open-source benchmarks for physical design tasks. E.g., IWLS 2005 benchmarks \cite{iwls} provide a repository of $85$ synthesized RTL netlists that were collected from various public resources \cite{opencores}. For discrete gate-sizing contest, ISPD 2012 \cite{ispd12} benchmarks annotate interconnect parasitics and timing constraints for 14 IWLS netlists.  ISPD 2015 benchmarks \cite{ispd15} provide a $65nm$ dataset that comprises eight designs with routing constraints and design rules, for the purpose of blockage-aware	detailed routing-driven placement task. For the multi-deck standard cell placement legalization problem, ICCAD 2017 \cite{iccad17} provides eight benchmarks in \texttt{LEF} and \texttt{DEF} format, along with soft placement constraints. They are derived from ISPD 2015 dataset. For the detailed routing contest of ISPD, ISPD 2018 benchmarks \cite{ispd18} provide $10$ test designs in $32nm$ and $45nm$ nodes, extracted from two real designs. It also defines design rules such as spacing tables, end-of-line spacing rules, cut spacing rules, min-area rules, and routing preference rules. For the macro-placement task, the ISPD02 IBM-MS mixed-size placement dataset \cite{ispd02} contains $18$ designs with both hard macros and standard cells. The TILOS-AI-Institute added four more designs to reproduce the results of the RL-based \textit{MacroPlacement} algorithm \cite{ct2021}. Though these efforts help in reproducibility, the scale of these benchmarks make it unusable to train modern ML models for physical design. 

\noindent \textbf{Floorplanning Benchmarks:}
For the floorplanning task, prior datasets GSRC \cite{gsrc} (\textit{n10, n30, n50, n100, n200,} and \textit{n300}), and MCNC \cite{mcnc} (\textit{ami33, ami49, apte, xp,} and \textit{xerox}) offer a standardized way to validate floorplan optimization algorithms. However, these benchmarks also suffer from being small scale, making them unusable for ML applications. In addition, the constraints used in these benchmarks do not capture many modern SoC floorplan constraints such as pre-placed constraints, boundary constraints, multi-instantiated partitions, and pin and net topologies.

\noindent \textbf{Synthetic Data in EDA:}
IP issues often make it difficult for commercial design companies to distribute their historical EDA data. 
Synthetic data offers a practical solution to this data paucity problem. However, for it to be usable, it needs to reflect the structural characteristics of real designs. Gupta et al. \cite{eyecharts} proposed benchmark circuits (called \textit{eyecharts}) of arbitrary size, to diagnose the weaknesses of existing gate-sizing algorithms, and to enable a systematic and quantitative comparison of sizing algorithms. 
In later works, Han et al. \cite{gt14} use artificial combinational paths to extract the  sign-off timing values for ground truths. The artificial circuits in their work are created by sweeping the number of stages, fan-outs and segments, and cell types in
timing paths. Kahng et al. \cite{cornerpred} also use artificial circuits to train a regression model, to predict timing values at unobserved corners. Recently, \texttt{PROBE2.0} \cite{probe2} proposed an artificial circuit with a mesh-like netlist topology as a place-and-route (P\&R) benchmark, for routability assessment. However, these artificial circuit timing paths do not reflect the distribution present in real circuits and, thus, do not capture the full complexity of real circuits. 
Kim et al. \cite{kim23} address the aforementioned issues and introduce an artificial netlist generator (ANG) framework for constructing authentic P\&R benchmarks suitable for ML. This framework generates gate-level netlists based on user-defined input parameters representing the topological attributes of realistic circuit. Our work shares a close affinity with the ANG framework, although we address the floorplanning task. To the best of our knowledge, we are the first to propose a large-scale dataset to enable ML techniques for floorplanning..

\section{Methodology}
\label{sec:methodology}

\subsection{Problem Formulation}

The \floorset \: benchmarks  specifically target the first step of the floorplanning task within SoC and sub-system hierarchies, aiming to identify optimal shapes and positions for sub-systems and partitions while adhering to certain placement constraints. 
The inputs to perform floorplanning task at SoC and sub-system hierarchies constitute:
\begin{itemize}
    \item area budgets of partitions or sub-systems.
    \item locations of external terminals of the system.
    \item netlist connectivity that define the connections between various components of the system.
    \item placement constraints that define pre-determined positions for some components of the system.
\end{itemize}

\begin{figure*}[!h]
    \centering
\includegraphics[scale=0.4]{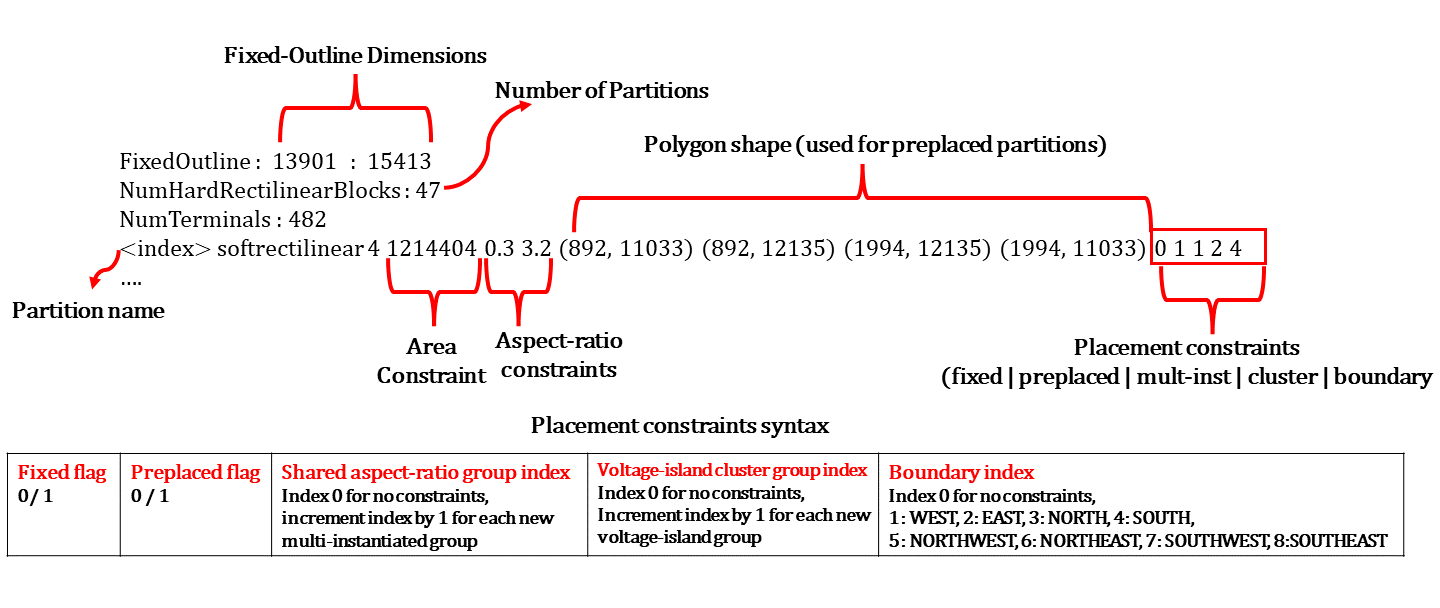}
\vspace{-0.3in}
\caption{The bookshelf \texttt{*.blocks} file is modified to include fixed-outline dimensions, area budgets, shape constraints (aspect ratio range) and placement constraints.}
    \label{fig:blocks_file}
\end{figure*}

\begin{figure}[!h]
    \centering
\includegraphics[scale=0.4]{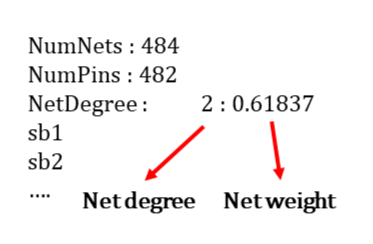}
\vspace{-0.2in}
\caption{The bookshelf \texttt{*.nets} file is modified to add net weights.}
    \label{fig:nets_file}
\end{figure}

The \texttt{bookshelf} format \cite{bookshelf} is a widely adopted, open-source, and standardized representation of VLSI design data. The components of a floorplanning problem are captured in the \texttt{*.blocks}, \texttt{*.nets} and \texttt{*.pl} file formats of the \texttt{bookshelf} format. As shown in Figure \ref{fig:blocks_file}, we modify the \texttt{*.blocks} to include component-wise placement constraints, area budgets, and fixed-outline dimensions. We include net weights in the \texttt{*.nets} file, shown in Figure \ref{fig:nets_file}. The \texttt{*.pl} file is unchanged and contains the $(x, y)$ locations of terminals. 

\subsection{Distribution of Real Layouts}

In this section, we outline the topological statistics derived from real industrial floorplans extracted from heterogeneous SoCs and their sub-systems, with configurations upto 100 partitions. These real floorplan layouts capture modern human-designed heterogeneous SoC and sub-system implementations in the industry. Due to the necessity of preserving the IP rights of these SoCs and their implementations, we are unable to disclose the exact layouts and their associated statistics. However, we provide the details of statistics that represent these realistic floorplan layouts.
We extract $10$ parameters, listed in Table \ref{tab:real_params}. These $10$ distributions allow us to systematically explore the design space and produce synthetic layouts that closely mimic real-world layouts. 

 The rationale for extracting these statistics is listed below:
\begin{itemize}
    \item $\mathit{A_{parts}}$: The aspect ratio range  defines the acceptable width-to-height ratio of a component (partition or sub-system based on the hierarchy). For rectilinear partitions, we derive the aspect ratio from the bounding box of the rectilinear polygon. This parameter guarantees that the generated synthetic layouts exhibit realistic shapes.
    
    \item $\mathit{R_{terms}^{parts}}$: The ratio of terminal count to partition count captures the proportion of terminals relative to partitions within a hierarchy. This offers a method to represent the proportional distribution of partitions and terminals in the floorplan netlists, without biasing towards the absolute scale of terminal count.

    \item $\big\{ \mathit{D_{parts}}, \mathit{D_{terms}}, \mathit{W_{parts}} \big\}$:  To quantify the complexity of connectivity structure in layouts at SoC and sub-system hierarchies, we use density (complement of sparsity) parameters of the connectivity matrices (or adjacency matrices).
    
    $\mathit{D_{parts}}$ and $\mathit{D_{terms}}$: These parameters capture the interconnection complexity of
    the floorplan netlists and provide a scale-independent mechanism to capture inter-partition and partition-terminal net connectivity. 
    It is common to have multiple net connections (up to few thousands) between a partition pair, and weighted nets capture the strength of such connections. 
    To account for weights on nets, we also extract the net-weight distribution,$\mathit{W_{parts}}$, as a function of net-length. The net-lengths are normalized to the fixed-outline's circumference, and net-weights are scaled to a range of 0-1. 

    \item $\mathit{E_{parts}}$: This captures the relative distribution of components (or partitions) with edge-affinity (or boundary affinity). These components typically represent the input-output (IO) constrained partitions or sub-systems.
    \item $\big \{ \mathit{N_{clusters}}, \mathit{C_{parts}} \big \}: $ These capture the number of voltage islands and the number of components in these island regions.
    
    \item $\mathit{P_{parts}}$: The percentage of pre-placed partitions capture the relative distribution of partitions with pre-placement constraints. These are hard constraints on the positions of partitions that are either derived from prior design knowledge or known-optimal decisions.
    \item $\mathit{M_{parts}}$: The percentage of multi-instantiation constraints represents the proportion of partitions enforcing shape-sharing constraints in modular design flows. E.g., in a hierarchy, if four instances of a CPU core exist, they all share the same shape and area.
\end{itemize}


\begin{table}[h]
\centering
\caption{Design parameters whose values are sampled from statistical distributions derived from real SoCs.}
\begin{adjustbox}{width=\columnwidth}
\begin{tabular}{|l|c|}\hline
\textbf{Parameter} & \textbf{Description}  \\    \hline \hline
$\mathit{A_{parts}}$ &  Aspect ratio ($ = \frac{W}{H}$) of a partition's bounding box\\ \hline
$\mathit{N_{terms}^{parts}}$   &  Number of terminals in a hierarchy relative to the partition count   \\ \hline
$\mathit{D_{parts}}$    &   Non-zero element percentage in the inter-partition connectivity matrix\\ \hline
$\mathit{W_{parts}}$     &  Inter-partition weight distribution as a function of distance \\ \hline

$\mathit{D_{terms}}$    & Non-zero element percentage in the terminal partition connectivity matrix  \\ \hline
$\mathit{E_{parts}}$   & Percentage of partitions with edge and corner constraints\\ \hline
$\mathit{C_{parts}}$  & Percentage of partitions with grouping (or clustering) constraints\\ \hline
$\mathit{N_{clusters}}$  & Number of clusters in a hierarchy\\ \hline
$\mathit{P_{parts}}$   & Percentage of partitions with preplaced positions\\ \hline
$\mathit{M_{parts}}$  & Percentage of partitions with multi-instantiation constraint\\ \hline

\end{tabular}
\end{adjustbox}
\label{tab:real_params}
\end{table}





\subsection{\floorset-Prime: Rectilinear Partitions }  

\begin{figure}[!h]
    \centering
\includegraphics[scale=0.35]{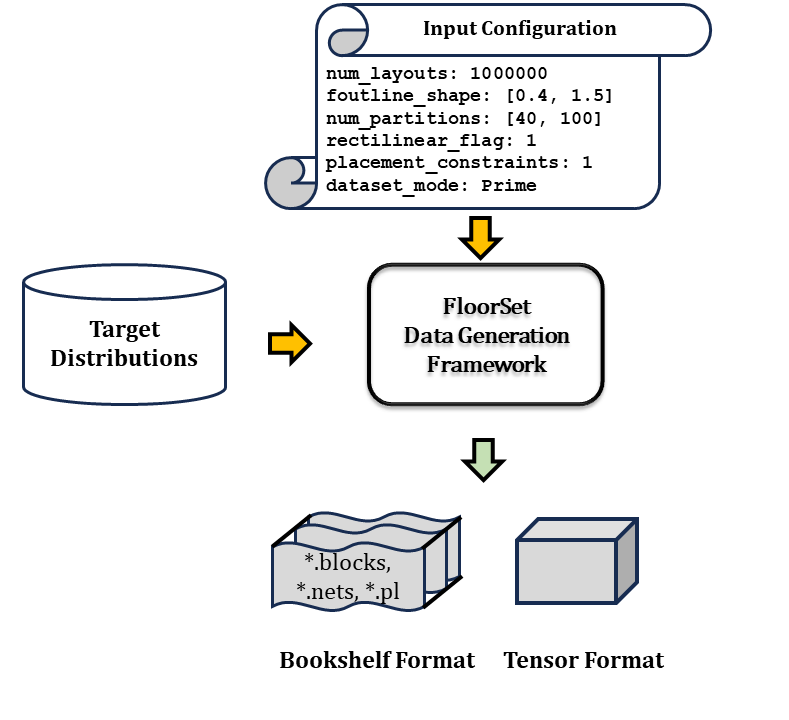}
\vspace{-0.3in}
\caption{The distribution of parameters (Table \ref{tab:real_params}) and the custom configuration file serve as  inputs for the data generation pipeline. The output layouts are formatted in the standard bookshelf format and Pytorch tensor format. }
    \label{fig:flow_fp2}
\end{figure}

\begin{figure*}[h]
    \centering
\includegraphics[scale=0.45]{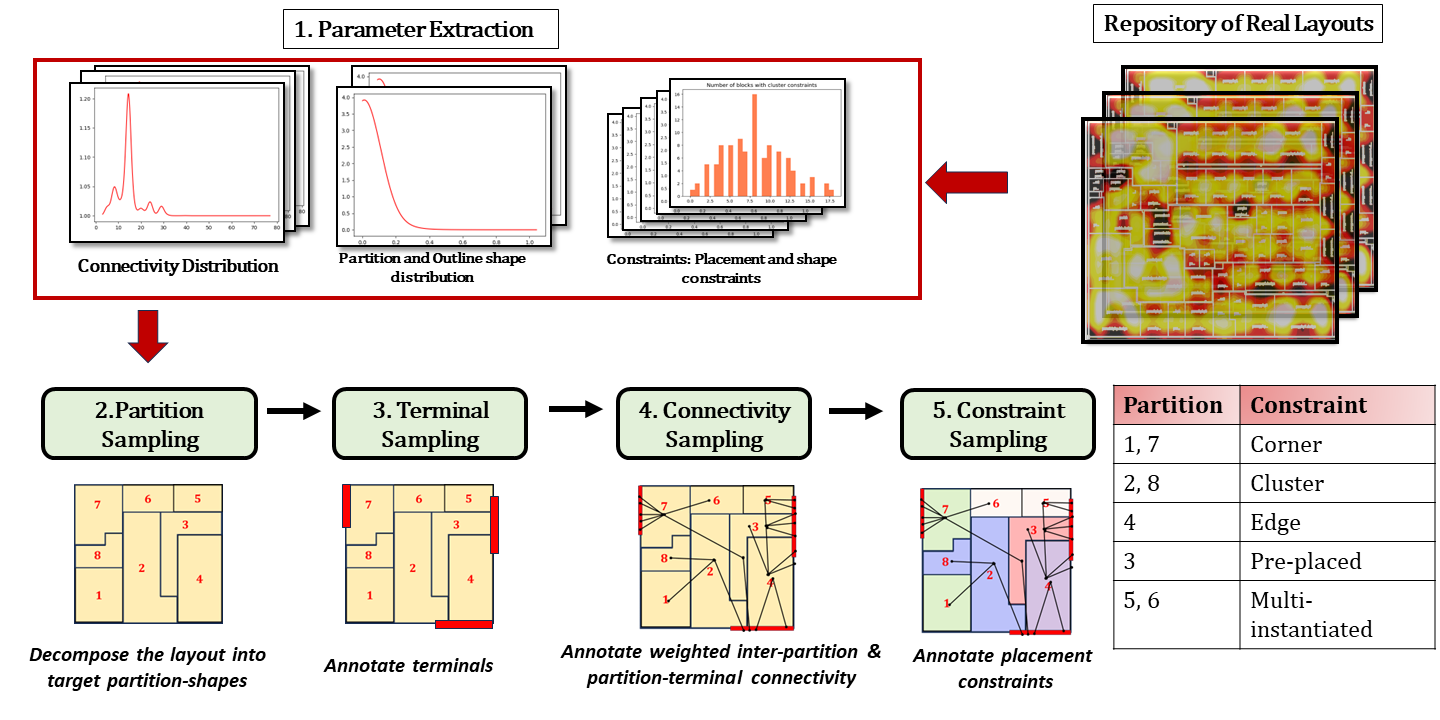}
\vspace{-0.2in}
\caption{Overview of the five-step \floorset \: data generation framework, illustrating the sequential processes involved in the methodology: 1. Collection and extraction of target layout distributions, 2. Partitioning shapes with the target area budgets, 3. Annotation of terminal locations, 4. Annotation of connectivity matrix (weighted), and 5. Annotation of placement.}
    \label{fig:flow_fp1}
\end{figure*}

In this section, we present the mechanism to generate synthetic layouts, using the floorplan statistics extracted from real layouts.
Figure \ref{fig:flow_fp2} provides an overview of the \floorset generation framework at a high-level. The circuit statistics extracted are utilized as an input for our data generation framework. 
In addition, the second input to the \floorset data generation pipeline is a custom input configuration file (shown in Figure \ref{fig:flow_fp2}), that contains the following settings: 
\begin{itemize}
    \item \texttt{num\_layouts}: to specify the number of floorplan layouts to be generated.
    \item \texttt{foutline\_shape}: to sweep the desired shape of the fixed-outline.
    \item \texttt{num\_partitions}: to sweep the target partition count.
    \item \texttt{rectilinear\_flag}: to enable or disable rectilinear partitions in the layout.
    \item \texttt{placement\_constraints\_flag}: to enable or disable annotation of hard placement constraints.
    \item \texttt{dataset\_mode flag}: to select between the available dataset pipelines (\floorset-Prime and \floorset-Lite).
\end{itemize}

The output of this data generation framework comprises layouts in \texttt{Bookshelf} and \texttt{Pytorch} tensor formats, representing diverse instances of floorplan problems. Expanding on the previous Figure \ref{fig:flow_fp2}, Figure \ref{fig:flow_fp1} explains the \floorset \: data generation framework using the following five important steps that are described below: Parameter Extraction, Partition Sampling, Terminal Sampling, Connectivity Sampling and Constraints Sampling.


The pseudocode of \floorset \: data generation pipeline is presented in Algorithm \ref{alg:datagen_main}. To recap, the inputs to the framework are target distributions extracted from real floorplan circuits, and input configuration file to sample the design parameters. The output of the Algorithm \ref{alg:datagen_main}
is a repository of synthetic floorplan layouts in the Bookshelf format and a tensor-based representation that is compatible with data loader libraries commonly used in ML applications.

\noindent \textbf{1. Parameter Extraction:} From a repository of several industrial floorplan layouts sourced from diverse SoC and subsystem implementations, we extract statistical metrics (listed in Table \ref{tab:real_params}) representative of the circuit characteristics inherent to  modern SoC and sub-system layouts. These metrics establish the target distribution from which we sample to produce synthetic floorplan layouts. To capture connectivity characteristics, we extract the density (complement of sparsity) distribution of inter-partition ($\mathit{D_{parts}}$) and partition-terminal ($\mathit{D_{terms}}$) adjacency matrices. In addition, we extract the relative proportion of terminal-count ($\mathit{N_{terms}^{parts}}$) to represent the scale of terminal count as a function of partition count in a hierarchy, and the the distribution of net-weights as a function of net-lengths ($\mathit{W_{parts}}$). Moreover, statistics regarding partition-specific aspect ratios ($\mathit{A_{parts}}$) and placement constraints ($\mathit{E_{parts}}$, $\mathit{C_{parts}}$, $\mathit{N_{clusters}}$, $\mathit{P_{parts}}$,
$\mathit{M_{parts}}$) are also extracted from real layouts.  The extracted statistics serve as the input to Algorithm \ref{alg:datagen_main}. 
Lines 1-2 initialize the "Layouts" database and index variables.  Line 3 is a loop for generating $n\_layouts$. We use tensor $\textbf{F}$ representation internally, to store the layout information (line 4).

\noindent \textbf{2. Partition Sampling:}  Following step 1, the next step (line 5) involves sweeping (or extracting) the desired fixed-outline dimensions from the input configuration file. Using this sampled outline, the synthetic layout undergoes decomposition into individual partitions or subsystems. To introduce variability, a mesh-layout (line 7 of Algorithm \ref{alg:datagen_main}) is created by employing a sequence of randomly sampled vertical and horizontal lines (line 6 of Algorithm \ref{alg:datagen_main}). These randomly positioned vertical and horizontal lines segment the fixed-outline layout into rectangles, forming the initial partition grid.  Subsequently, adjacent polygons are merged, to create partition shapes that match the desired target distribution of partition shapes. To ensure sufficient room for the creation of desired shapes and area budgets, we start with a large partition count ($4\times - 6\times$ of the desired partition count) on the initial grid (line 6 of of Algorithm \ref{alg:datagen_main}). The randomized selection of merging candidates and randomized initial grid creation offer a systematic approach to sampling numerous potential fully-abutted divisions of the fixed-outline layout, exposing complex but realistic use-cases.  The details of the merging process is elaborated in Algorithm \ref{alg:datagen_merge}. As shown in line 2 and line 9 of Algorithm \ref{alg:datagen_merge}, polygons on the layout are iteratively merged until the desired shapes and partition count are achieved. If \textbf{\texttt{rectilinear\_flag} = 0} (line 4), the merging operation ensures that generated partitions are always rectangles. Conversely, when the flag is set to 1, there are no restrictions on the rectilinearity of the shapes.  Each merging operation is committed if the merged partition operation improves the alignment between the current shape distribution and the target shape distribution (lines 7 and 8).

\begin{figure}[!h]
    \centering
\includegraphics[scale=0.3]{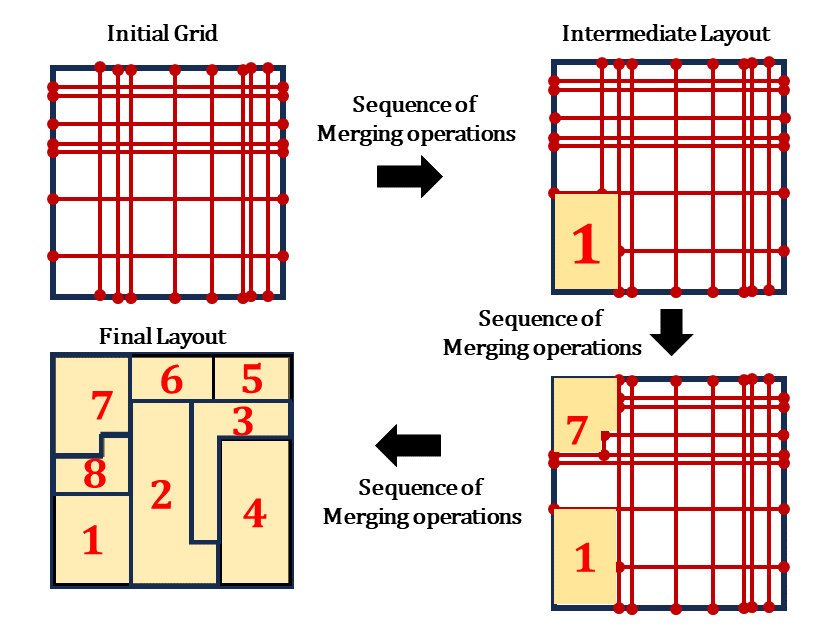}
\vspace{-0.2in}
\caption{A sequence of merging operations, to generate the channel-less layout with desired shapes and area budgets. }
    \label{fig:merging_flow}
\end{figure}

\begin{figure}[!h]
    \centering
\includegraphics[scale=0.3]{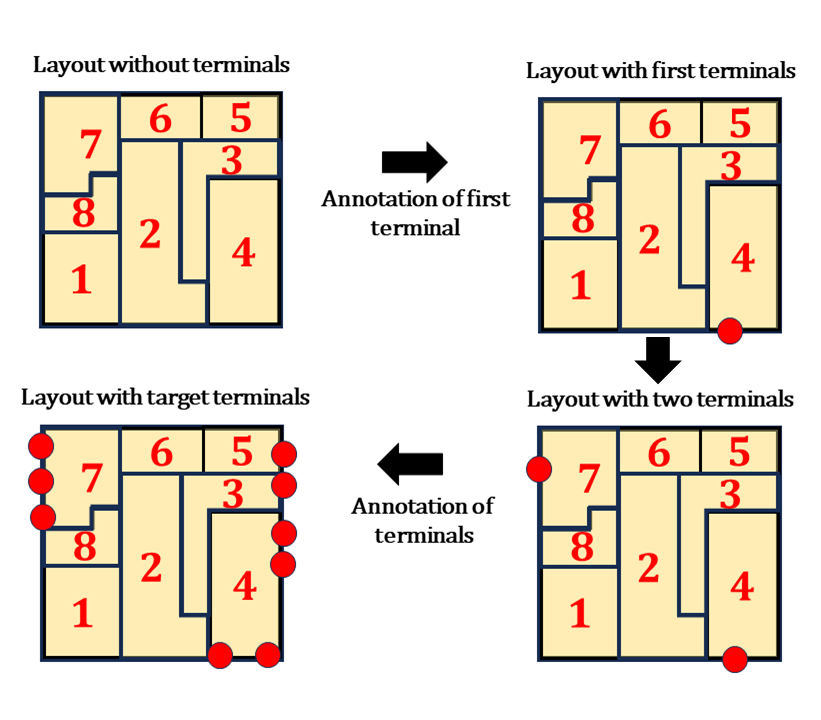}
\vspace{-0.2in}
\caption{A sequence of terminal-annotation operations, to place terminals on the layout from Step 1. }
    \label{fig:terminal_flow}
\end{figure}

\noindent \textbf{3. Terminal Sampling:} After the creation of the fully-abutted layout, we annotate the terminal positions on the outline (line 9 of Algorithm \ref{alg:datagen_main}). Illustrated in Figure \ref{fig:terminal_flow} and Algorithm \ref{alg:datagen_terms}, this terminal annotation process involves first placing a terminal at a random location along the outline (line 3 of Algorithm \ref{alg:datagen_terms}). Subsequently, other terminals are incrementally placed (line 6 of Algorithm \ref{alg:datagen_terms}), while adhering to the normalized-pitch considerations (line 8 of Algorithm \ref{alg:datagen_terms}) and meeting the total terminal-count quota.

\begin{figure}[!h]
    \centering
\includegraphics[scale=0.3]{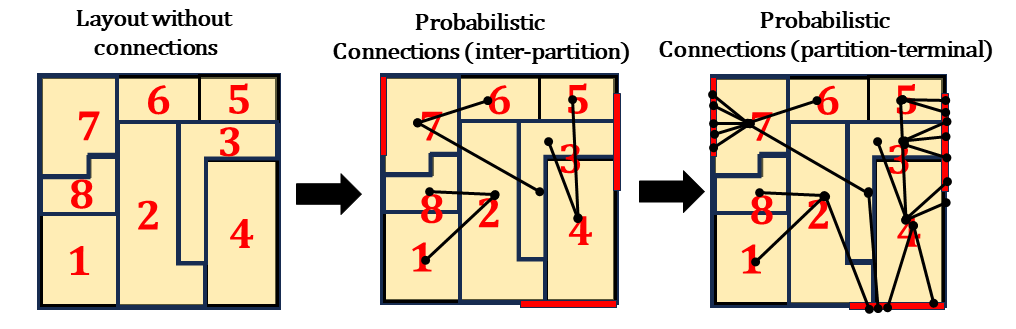}
\vspace{-0.2in}
\caption{Using the layouts with partition and terminals placed, connections are added using probabilities proportional to the closeness (Manhattan distance) of components. }
    \label{fig:connect_flow}
\end{figure}

\noindent \textbf{4. Connectivity Sampling:} Following the  placement of desired partition shapes (or sub-systems), and terminal locations, we create the connectivity between partitions and terminals (line 10 of Algorithm \ref{alg:datagen_nets}). The details of connectivity annotation are explained in Algorithm \ref{alg:datagen_nets}.
While decomposing the fixed outline into individual partitions implicitly ensures area-optimal layouts, a optimal-by-construct annotation of connectivity is necessary to generate an optimal or near-optimal connectivity matrix. To achieve this, we utilize the pair-wise distance matrix among all partition pairs, as well as partition and terminal combinations (lines 4-5 of Algorithm \ref{alg:datagen_nets}). The inverse of the distance matrix yields the "similarity matrix," which denotes the proximity of modules in $2$D Manhattan space (lines 6-7 of Algorithm \ref{alg:datagen_nets}).  Since the optimal layouts tend to place heavily connected partitions closer, annotating connections using the similarity matrix as the probabilities (line 8-9 of Algorithm \ref{alg:datagen_nets})  offers a rational approach to ensuring the near-optimality of generated connections. In addition, we also sample inter-partition net-weights from the target distribution that captures net-weights as a function of inter-partition distance. using the similarity matrix.  Figure \ref{fig:connect_flow} visually explain the annotation of weighted inter-partition nets and partition-terminal nets, that meet the target sparsity for inter-partition and partition-net connections. 

\begin{figure}[!h]
    \centering
\includegraphics[scale=0.3]{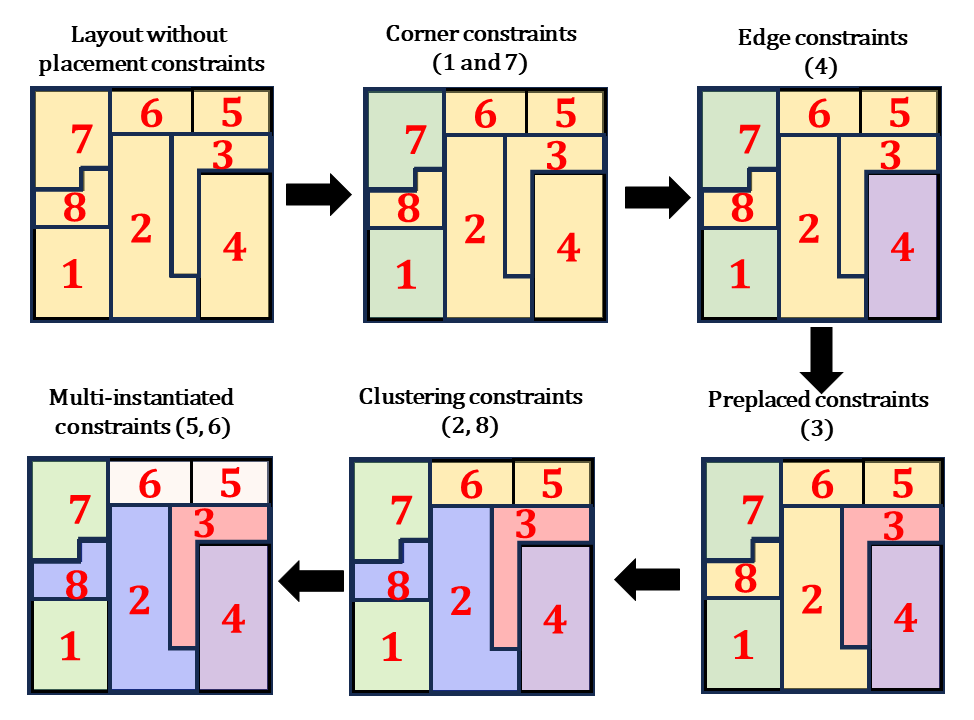}
\vspace{-0.2in}
\caption{Using the layouts with optimal shapes and locations, placement constraints are annotated. We also use connectivity from Step 4, to determine edge affinity. }
    \label{fig:constraint_flow}
\end{figure}

\noindent \textbf{5. Constraint Sampling:}
The output from Step 4 embodies the circuit characteristics of real layouts in terms of near-optimal connectivity and optimal area. Line 11 of Algorithm \ref{alg:datagen_main} is the constraint annotation step, where  the desired placement constraints are annotated on the area and wire-length optimal layouts. The resulting layout manifests as a hard-constrained floorplanning instance with optimal area and nearly optimal wire-length characteristics. Further details on constraint annotation are provided in Algorithm \ref{alg:datagen_constraints}. Utilizing the connectivity information and the optimal positions of individual partitions, boundary constraints such as corner partitions and edge partitions are annotated (line 7 of Algorithm \ref{alg:datagen_constraints}). In real circuits, it is commonly observed that the majority of pre-placed partitions reside towards the periphery, and therefore we annotate pre-placed constraint, with a higher affinity for partitions that are placed on the periphery of the layout (line 8 of Algorithm \ref{alg:datagen_constraints}). Line 9 of Algorithm \ref{alg:datagen_constraints} annotates the clustering constraint on a subset of partitions that are strongly connected (physical-adjacency requirement). Finally, multi-instantiated constraints are enforced on a subset of partitions that share identical shape and area values (line 10 of Algorithm \ref{alg:datagen_constraints}). Figure \ref{fig:constraint_flow} illustrates the process of constraint annotation. \\

\begin{algorithm}
\caption{\floorset-Prime: Generating optimal layouts with constraint annotation}
\begin{algorithmic}[1]
\REQUIRE Input Configuration: $I_{config}$ = num\_layouts,  foutline\_shape, num\_partitions,  rectilinear\_flag, placement\_constraints, dataset\_mode = Prime \\ Target distribution: $(\mathit{A_{parts}}, \mathit{N_{terms}^{parts}}, \mathit{D_{parts}}, \mathit{W_{parts}}, \mathit{D_{terms}},$ \\
$\mathit{E_{parts}}, \mathit{C_{parts}}, 
\mathit{N_{clusters}},
\mathit{P_{parts}},
\mathit{M_{parts}})$

\STATE $Layouts = \emptyset$
\STATE $index = 0$
\WHILE{$index < num\_layouts $ }
\STATE $\textbf{F} = \emptyset$ \hfill \emph{Internal geometric representation of the layout}
\STATE $W, H, n\_parts \leftarrow sample(foutline\_shape, num\_partitions)$ \\
\hfill \emph{Fixed-outline dimensions \& number of partitions}

\STATE $n\_grids \leftarrow random(4, 6)*n\_parts$ \hfill \emph{Initial partition count}
\STATE $\textbf{F} \leftarrow createMESH(n\_grids, W, H)$ \hfill \\
\hfill \emph{Random partitioning of the layout in to n\_grids}
\STATE $\textbf{F} \leftarrow mergePARTITION(n\_parts,\mathit{A_{parts}}, rectilinear\_flag, \textbf{F})$ \hfill \\ \hfill \emph{Merging of partitions to get target shapes}
\STATE $\textbf{F} \leftarrow annotateTERMINALS(\mathit{N_{terms}^{parts}}, n\_parts, term\_pitch \textbf{F})$  \\
\hfill \emph{Annotate terminals on the outline of the layout} \\
\STATE $\textbf{F} \leftarrow 
 annotateNETS(\mathit{D_{parts}}, \mathit{D_{terms}}, \mathit{W_{parts}}, n\_parts,  \textbf{F})$ \hfill \\ \hfill \emph{Create connections among partitions and terminals}
 \IF {placement\_constraints == 1}
\STATE $\textbf{F} \leftarrow annotateCONSTRAINTS(\mathit{E_{parts}},  \mathit{C_{parts}}, \mathit{N_{clusters}},
\mathit{P_{parts}},$ \\
\hfill $\mathit{M_{parts}},\textbf{F})$ \\ \hfill 
\emph{Annotate placement constraints on the partitions}
\ENDIF

\STATE $index += 1$
\STATE $blocks, nets, pl \leftarrow convertBOOKSHELF(\textbf{F})$
\STATE $Layouts[index] = [blocks, nets, pl, \textbf{F}]$

\ENDWHILE
\end{algorithmic}
\label{alg:datagen_main}
\end{algorithm}

\begin{algorithm}
\caption{Merging of partitions: mergePARTITION}
\begin{algorithmic}[1]
\REQUIRE $n\_parts, \mathit{A_{parts}}, rectilinear\_flag, \textbf{F}$
\STATE $part\_count = countPolygons(\textbf{F})$
\WHILE{$part\_count > n\_parts$}
    \STATE $\textbf{F\_new} \leftarrow removeRandomEdge(\textbf{F})$
    \IF{$rectilinear\_flag == 0$}
    \IF{checkRectilinear(\textbf{F\_new})}
            \STATE continue
        \ENDIF
    \ENDIF  
    \IF {wassertsteinDist(\textbf{F\_new}, \textbf{T}) < wassertsteinDist(\textbf{F}, \textbf{T})} 
        \STATE $\textbf{F} \leftarrow \textbf{F\_new}$
        \STATE $part\_count = countPolygons(\textbf{F})$
    \ENDIF
\ENDWHILE
\STATE return \textbf{F}
\end{algorithmic}
\label{alg:datagen_merge}
\end{algorithm}

\begin{algorithm}
\caption{Annotation of terminals: annotateTERMINALS}
\begin{algorithmic}[1]
\REQUIRE $\mathit{N_{terms}^{parts}}, n\_parts,  \textbf{F}$
\STATE $n\_terms =  int(n\_parts*\mathit{N_{terms}^{parts}})$
\STATE $term\_count = 0$
\STATE $W, H \leftarrow extractOutline(\textbf{F})$
\STATE $term\_pitch = 2*(W+H)/n\_terms$
\STATE $tx, ty \leftarrow randomPointOnOutline(\textbf{F})$
\STATE $term\_count += 1$
\STATE $TermList[term\_count] = [tx, ty]$
\WHILE{$term\_count < n\_terms$}
    \STATE $tx, ty \leftarrow randomPointOnOutline(\textbf{F})$
    \IF {$minDist(TermList, [tx, ty]) \geq term\_pitch$} 
        \STATE $term\_count += 1$
        \STATE $TermList[term\_count] = [tx, ty]$
    \ENDIF
\ENDWHILE
\STATE $\textbf{F} \leftarrow addTermLocs(TermList)$
\STATE return \textbf{F}
\end{algorithmic}
\label{alg:datagen_terms}
\end{algorithm}

\begin{algorithm}
\caption{Annotation of Connectivity: annotateNETS}
\begin{algorithmic}[1]
\REQUIRE $\mathit{D_{parts}},  
\mathit{D_{terms}}, \mathit{W_{parts}},
\textbf{F}$
\STATE $n\_b2b\_nets =  int(n\_parts*\mathit{D_{parts}})$\\
\hfill \emph{Number of inter-partition nets}\\

\STATE $n\_t2b\_nets =  n\_nets *\mathit{D_{terms}}$ \\
\hfill \emph{Number of partition-terminal nets}\\
\STATE $PDist \leftarrow pairwiseB2BDistance(\textbf{F})$
\STATE $TDist \leftarrow pairwiseT2BDistance(\textbf{F})$\\
\hfill \emph{Extract Manhattan distance}
\STATE $PSim \leftarrow 1 - Normalize(PDist) $
\STATE $TSim \leftarrow 1 - Normalize(TDist) $\\
\hfill \emph{Probabilities measured as inverse of distance}
\STATE $b2bConnectivity \leftarrow Sample(\textbf{F}, size = n\_b2b\_nets, p = PSim)$
\STATE $b2bweights \leftarrow Sample(\mathit{W_{parts}}, size = n\_b2b\_nets, p = PSim)$
\STATE $t2bConnectivity \leftarrow Sample(\textbf{F}, size = n\_p2b\_nets, p = TSim)$
\STATE $\textbf{F} \leftarrow addNets(b2bConnectivity, b2bweights, t2bConnectivity)$\\
\hfill \emph{Sample net-weights and net-connections using probabilities}
\STATE return \textbf{F}
\end{algorithmic}
\label{alg:datagen_nets}
\end{algorithm}

\begin{algorithm}
\caption{Annotation of Placement Constraints: annotateCONSTRAINTS}
\begin{algorithmic}[1]
\REQUIRE $\mathit{E_{parts}}, \mathit{C_{parts}, \mathit{N_{clusters}}, \mathit{P_{parts}}, \mathit{M_{parts}}}  \textbf{F}$
\STATE $n\_boundary\_parts =  int(n\_parts*\mathit{E_{parts}})$
\STATE $n\_clustered\_parts =  int(n\_parts*\mathit{C_{parts}})$
\STATE $n\_preplaced\_parts =  int(n\_parts*\mathit{P_{parts}})$
\STATE $n\_multi\_inst\_parts =  int(n\_parts*\mathit{M_{parts}})$
\STATE $eDist \leftarrow inverse edgeDistances(\textbf{F})$
\STATE $cDist \leftarrow connectivity(\textbf{F})$
\STATE $\textbf{F\_e} \leftarrow getEdgeParts(\textbf{F})$
\STATE $bParts \leftarrow Sample(\textbf{F\_e}, size = n\_boundary\_parts, p = cDist)$
\STATE $pParts \leftarrow Sample(\textbf{F}, size = n\_preplaced\_parts, p = eDist)$

\STATE $cParts \leftarrow deriveClusters(\textbf{F}, n\_clustered\_parts, \mathit{N_{clusters}})$

\STATE $mParts \leftarrow deriveMultiInst(\textbf{F}, n\_multi\_inst\_parts)$

\STATE $\textbf{F} \leftarrow addConstraints(bParts, cParts, pParts, mParts)$
\STATE return \textbf{F}
\end{algorithmic}
\label{alg:datagen_constraints}
\end{algorithm}

\noindent \textbf{Special case of \floorset-Prime (\textbf{rectilinear\_flag = 0):}} Although we observe that partition shapes or sub-system shapes in real circuits are rectilinear, the current literature is far from handing arbitrarily rectilinear partitions under hard constraints. Therefore, we provide a mechanism to generate a simpler case of the\floorset-Prime dataset by setting rectilinear\_flag to $0$. The layouts thus generated will be fully abutted rectangular shapes. The near-optimal connectivity and annotation of hard constraints remain unchanged.

\subsection{\floorset-Lite: Rectangular Partitions} 
Fully abutted, rectilinear shapes of partitions in a SoC or a sub-system typically emerge in the later stages of floorplanning. 
Such rectilinear shapes truly capture the complexity of industrial floorplanning formulation.  However, in the initial design exploration stages, designers commonly begin with rectangular partitions, prioritizing finding optimal locations and early design goals (e.g., timing budgets and area requirements). At this stage, the assumption of rectangular partitions often results in the formation of channels or white-spaces. This use-case is also an important problem for design-space exploration, although less challenging than the fully-rectilinear counterpart. Therefore, we introduce an additional dataset, \floorset-Lite, to reflect such scenarios. The pseudocode of \floorset-Lite pipeline is shown in Algorithm \ref{alg:datagen_lite}. While \floorset-Prime decomposes the fixed-outline layout into fully-abutted rectilinear partitions, \floorset-Lite allows for gaps or whites-paces on the layout, to accommodate the rectangular assumption of partitions. As a result, we employ a slight change in the data generation pipeline to generate layouts with white-space while maintaining near-optimal characteristics of area and wire-length. It is well established that heuristic search algorithms such as Simulated Annealing (SA) excel at the unconstrained floorplanning problem \cite{adya01, adya03}, particularly when the joint objective of wire-length and area is removed. When the focus shifts solely to minimizing white-space, SA demonstrates exceptional performance. Therefore, as shown in line 6 of Algorithm \ref{alg:datagen_lite}, we use SA to generate a packing solution for rectangular partitions while respecting the provided fixed-outline aspect ratio. This solution is near-optimal by construction in terms of area, with some white-space (primarily because of the rectangular assumption). As shown in lines 7-11, the rest of the process (terminal annotation, connectivity annotation and placement-constraint annotation) is identical to that of \floorset-Prime. Similar to the \floorset-Prime flow, \floorset-Lite uses the configuration file (Figure \ref{fig:flow_fp2}) to generate 1M training samples and 100 validation test cases.

\begin{algorithm}
\caption{\floorset-Lite: Generating optimal layouts with constraint annotation}
\begin{algorithmic}[1]
\REQUIRE Input Configuration: $I_{config}$ = num\_layouts,  foutline\_shape, num\_partitions,  rectilinear\_flag, placement\_constraints, dataset\_mode = Prime \\ Target distribution: $(\mathit{A_{parts}}, \mathit{N_{terms}^{parts}}, \mathit{D_{parts}}, \mathit{W_{parts}}, \mathit{D_{terms}},$ \\
$\mathit{E_{parts}}, \mathit{C_{parts}}, 
\mathit{N_{clusters}},
\mathit{P_{parts}},
\mathit{M_{parts}})$

\STATE $Layouts = \emptyset$
\STATE $index = 0$
\WHILE{$index < num\_layouts $ }
\STATE $\textbf{F} = \emptyset$ \hfill \emph{Internal geometric representation of the layout}
\STATE $W, H, n\_parts \leftarrow sample(foutline\_shape, num\_partitions)$ \\
\hfill \emph{Fixed-outline dimensions \& number of partitions}

\STATE  $\textbf{F} \leftarrow runSA(W, H, n\_parts, \mathit{A_{parts}})$ \hfill \\
\hfill \emph{Run Parquet (Simulated Annealing) for area optimization} \\

\STATE $\textbf{F} \leftarrow annotateTERMINALS(\mathit{N_{terms}^{parts}}, n\_parts, term\_pitch \textbf{F})$  \\
\STATE $\textbf{F} \leftarrow 
 annotateNETS(\mathit{D_{parts}}, \mathit{D_{terms}}, \mathit{W_{parts}}, n\_parts,  \textbf{F})$ \hfill \\ 
 \IF {placement\_constraints == 1}
\STATE $\textbf{F} \leftarrow annotateCONSTRAINTS(\mathit{E_{parts}},  \mathit{C_{parts}}, \mathit{N_{clusters}},
\mathit{P_{parts}},$ \\
\hfill $\mathit{M_{parts}},\textbf{F})$ \\ 
\ENDIF

\STATE $index += 1$
\STATE $blocks, nets, pl \leftarrow convertBOOKSHELF(\textbf{F})$
\STATE $Layouts[index] = [blocks, nets, pl, \textbf{F}]$

\ENDWHILE
\end{algorithmic}
\label{alg:datagen_lite}
\end{algorithm}



\section{Experiments}
\label{sec:experiments}

\subsection{Analysis of FloorSet}

While the $1$M synthetic training layouts in \floorset-Prime and \floorset-Lite are intended for training ML models, we reserve the $100$ test cases in each dataset to be used as a standard benchmark. In each dataset, the following labels serve as performance metrics: (a) area, (b) weighted wire-lengths (inter-partition and partition-to-terminal), and (c) violation count for each of the five placement constraints. E.g., if tensor \textbf{F} represents the validation dataset, one can retrieve the metric list of test-case $i$ using $\textbf{F[$i$]["labels"]}$. The metric list has three components, in a list format [area, inter-partition wire-length, terminal-to-partition wire-length]. Since the dataset is devoid of constraint violations by construction, we do not explicitly store the violation count. 

\begin{figure}[!h]
    \centering
\includegraphics[scale=0.32]{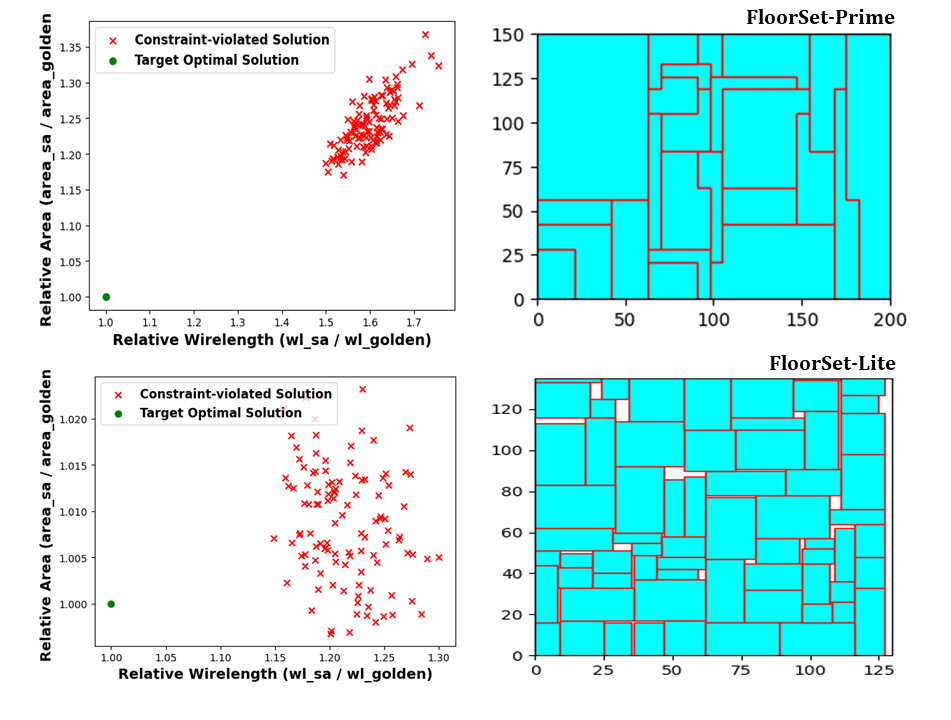}
\vspace{-0.2in}
\caption{Scatter plots (left) and the corresponding constraint annotated optimal layouts (right) indicating the optimality gap and violations of hard constraints, when applying classical SA on a test cases picked from the validation set of  \floorset-Prime (top) and \floorset-Lite (bottom). On \floorset-Lite, SA is able to find more area-optimal placements compared to \floorset-Prime - consistent with the former's simplified rectangular partitions. However, almost all solutions discovered by SA on both datasets have constraint violations.}
\label{fig:floorsetp_sa}
\end{figure}

To underscore the complexity of \floorset, we run classical SA \cite{btree06} with a constraint-penalty on the \floorset \: validation set. Figure \ref{fig:floorsetp_sa} presents scatter plots (left) for one data-point from the validation sets of \floorset-Prime (top) and \floorset-Lite (bottom). The relative area and wire-length are derived using the corresponding metrics from \floorset \ as the baseline; the golden layout on this relative scale is indicated by the green dot at ($1$, $1$). For \floorset-Prime, we observe that the solutions are significantly sub-optimal in terms of wire-length and area. Crucially, all solutions exhibit constraint violations. While solutions obtained on \floorset-Lite's problem are closer to the optimal in terms of area, minimizing wire-length and fixing constraint violations remain a big challenge.  The desired behavior of any future baseline is to approach the green dots, while respecting the hard-constraints. 
Figure \ref{fig:floorsetp_sa} also shows the layout images of the corresponding golden data. The top-right figure shows an example 25-partition \floorset-Prime layout (corresponding to the green dot on top-left figure). The bottom-right figure shows an example 65-partition \floorset-Lite layout (corresponding to the green dot on bottom-left figure). The synthetic layout of \floorset-Prime (top-right) closely resembles partition shapes in real-word layouts.

\subsection{Distributions of \floorset \ Characteristics}
\begin{figure}[!h]
    \centering
\includegraphics[scale=0.32]{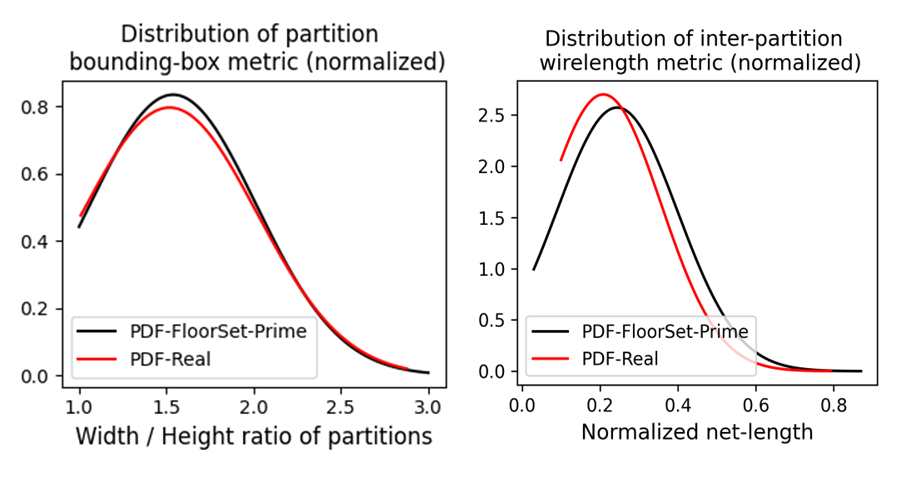}
\vspace{-0.2in}
\caption{Comparison between distributions of partition characteristics in \floorset \: vs real designs. Distributions of aspect ratio (left) and inter-partition wire-length (right) in \floorset \: closely match that of real SoC designs.}
\label{fig:pdf}
\end{figure}

While Section 4.1 underscores the complexity of \floorset-Prime benchmarks and highlights the sub-optimal performance of existing methods, it remains essential to establish the resemblance between these complex synthetic layouts and real-world counterparts in terms of their circuit characteristics. Figure \ref{fig:pdf} compares the probability density function (PDF) of synthetic and real layouts using their shape and connectivity metrics. The left plot depicts the partition-level shape distribution by plotting the PDF with respect to the aspect ratios of individual partitions. The proximity of real PDF (red) and the \floorset-Prime PDF (black) is a clear indication of the resemblance between the shapes of synthetic and real layouts. Similarly, the plot on the right shows the PDF of inter-partition wire-length. The normalization of wire-length allows for a direct comparison. The closeness of the red and black curves illustrates that the synthetic layouts closely reflect the inherent connectivity distribution observed in real-world layouts. 

\section{Conclusions}
\label{sec:conclusions}

This paper introduces FloorSet, a VLSI floorplanning dataset that reflects the complexities and constraints of modern SoC design. Crucially, it provides large-scale training data that is necessary to develop modern ML techniques to solve floorplanning problems. \floorset-Lite, with near-optimal packing and wire-length for rectangular partitions, represents early stages of design. \floorset-Prime, with the added complexity of fully-abutted rectilinear blocks, represents the final stages of design. \floorset \ is constructed by sampling outline and wire-length values from distributions derived from real, commercial SoC designs. We hope that \floorset \ spurs novel solutions in the area of complex combinatorial optimization that significantly advance the field of EDA.

\section{Acknowledgements}
\label{sec:acknowledgments}
We thank Miaomiao Ma (Intel), Pei Chun Ch'ng (Intel), Olena Zhu (Intel) and Fadi Aboud (Intel) for providing us with insights into the real-world constraints that are often overlooked in academic benchmarks. Their expertise and collaboration have been critical to the development of \floorset.

\printbibliography

@ARTICLE{mlcad22,
  author={Rapp, Martin and Amrouch, Hussam and Lin, Yibo and Yu, Bei and Pan, David Z. and Wolf, Marilyn and Henkel, Jörg},
  journal={IEEE Transactions on Computer-Aided Design of Integrated Circuits and Systems}, 
  title={MLCAD: A Survey of Research in Machine Learning for CAD Keynote Paper}, 
  year={2022},
  volume={41},
  number={10},
  pages={3162-3181}}

@ARTICLE{mleda23,
  author={Kahng, Andrew B.},
  journal={IEEE Design \& Test}, 
  title={Machine Learning for CAD/EDA: The Road Ahead}, 
  year={2023},
  volume={40},
  number={1},
  pages={8-16}}

@unpublished{dpan21,
title= {Closing the Virtuous Cycle of
AI for IC and IC for AI},
author = {David Z. Pan},
year = {2021},
note= {The Council on Electronic Design Automation (CEDA), IEEE},
URL= {https://ieee-ceda.org/presentation/webinar/closing-virtuous-cycle-ai-ic-and-ic-ai},
}

@INPROCEEDINGS{learnfp20,
  author={He, Zhuolun and Ma, Yuzhe and Zhang, Lu and Liao, Peiyu and Wong, Ngai and Yu, Bei and Wong, Martin D.F.},
  booktitle={2020 IEEE 38th International Conference on Computer Design (ICCD)}, 
  title={Learn to Floorplan through Acquisition of Effective Local Search Heuristics}, 
  year={2020},
  volume={},
  number={},
  pages={324-331}}

@inproceedings{binpacking21,
  author    = {Hang Zhao and
               Qijin She and
               Chenyang Zhu and
               Yin Yang and
               Kai Xu},
  title     = {Online 3D Bin Packing with Constrained Deep Reinforcement Learning},
  booktitle = {Thirty-Fifth {AAAI} Conference on Artificial Intelligence, {AAAI}
               2021},
  pages     = {741--749},
  publisher = {{AAAI} Press},
  year      = {2021}

}

@inproceedings{timinggnn22,
author = {Guo, Zizheng and Liu, Mingjie and Gu, Jiaqi and Zhang, Shuhan and Pan, David Z. and Lin, Yibo},
title = {A timing engine inspired graph neural network model for pre-routing slack prediction},
year = {2022},
isbn = {9781450391429},
publisher = {Association for Computing Machinery},
address = {New York, NY, USA},
booktitle = {Proceedings of the 59th ACM/IEEE Design Automation Conference},
pages = {1207–1212},
numpages = {6},
location = {San Francisco, California},
series = {DAC '22}
}

@InProceedings{chipformer23,
  title = 	 {{C}hi{PF}ormer: Transferable Chip Placement via Offline Decision Transformer},
  author =       {Lai, Yao and Liu, Jinxin and Tang, Zhentao and Wang, Bin and Hao, Jianye and Luo, Ping},

  pages = 	 {18346--18364},
  year = 	 {2023},
  volume = 	 {202},
  series = 	 {Proceedings of Machine Learning Research},
  month = 	 {23--29 Jul},
  publisher =    {PMLR}
}

@misc{chipnemo24,
      title={ChipNeMo: Domain-Adapted LLMs for Chip Design}, 
      author={Mingjie Liu and Teodor-Dumitru Ene and Robert Kirby and Chris Cheng and Nathaniel Pinckney and Rongjian Liang and Jonah Alben and Himyanshu Anand and Sanmitra Banerjee and Ismet Bayraktaroglu and Bonita Bhaskaran and Bryan Catanzaro and Arjun Chaudhuri and Sharon Clay and Bill Dally and Laura Dang and Parikshit Deshpande and Siddhanth Dhodhi and Sameer Halepete and Eric Hill and Jiashang Hu and Sumit Jain and Ankit Jindal and Brucek Khailany and George Kokai and Kishor Kunal and Xiaowei Li and Charley Lind and Hao Liu and Stuart Oberman and Sujeet Omar and Ghasem Pasandi and Sreedhar Pratty and Jonathan Raiman and Ambar Sarkar and Zhengjiang Shao and Hanfei Sun and Pratik P Suthar and Varun Tej and Walker Turner and Kaizhe Xu and Haoxing Ren},
      year={2024},
      eprint={2311.00176},
      archivePrefix={arXiv},
      primaryClass={cs.CL}
}

@INPROCEEDINGS{chipchat23,
  author={Blocklove, Jason and Garg, Siddharth and Karri, Ramesh and Pearce, Hammond},
  booktitle={2023 ACM/IEEE 5th Workshop on Machine Learning for CAD (MLCAD)}, 
  title={Chip-Chat: Challenges and Opportunities in Conversational Hardware Design}, 
  year={2023},
  volume={},
  number={},
  pages={1-6}}

@INPROCEEDINGS{agnesia20,
  author={Agnesina, Anthony and Chang, Kyungwook and Lim, Sung Kyu},
  booktitle={ICCAD '20}, 
  title={VLSI Placement Parameter Optimization using Deep Reinforcement Learning}, 
  year={2020},
  volume={},
  number={},
  pages={1-9},
  doi={}}

@INPROCEEDINGS{rlsizer21,
  author={Lu, Yi-Chen and Nath, Siddhartha and Khandelwal, Vishal and Lim, Sung Kyu},
  booktitle={DAC '21}, 
  title={RL-Sizer: VLSI Gate Sizing for Timing Optimization using Deep Reinforcement Learning}, 
  year={2021},
  volume={},
  number={},
  pages={733-738}}

@INPROCEEDINGS{pdn20,
  author={Chhabria, Vidya A. and Kahng, Andrew B. and Kim, Minsoo and Mallappa, Uday and Sapatnekar, Sachin S. and Xu, Bangqi},
  booktitle={ASP-DAC}, 
  title={Template-based PDN Synthesis in Floorplan and Placement Using Classifier and CNN Techniques}, 
  year={2020},
  volume={},
  number={},
  pages={44-49}}

@inproceedings{drc20,
author = {Liang, Rongjian and Xiang, Hua and Pandey, Diwesh and Reddy, Lakshmi and Ramji, Shyam and Nam, Gi-Joon and Hu, Jiang},
title = {DRC Hotspot Prediction at Sub-10nm Process Nodes Using Customized Convolutional Network},
year = {2020},
isbn = {9781450370912},
publisher = {Association for Computing Machinery},
address = {New York, NY, USA},

booktitle = {Proceedings of the 2020 International Symposium on Physical Design},
pages = {135–142},
numpages = {8},
series = {ISPD '20}
}

@article{dagsizer23,
author = {Cheng, Chung-Kuan and Holtz, Chester and Kahng, Andrew B. and Lin, Bill and Mallappa, Uday},
title = {DAGSizer: A Directed Graph Convolutional Network Approach to Discrete Gate Sizing of VLSI Graphs},
year = {2023},
issue_date = {July 2023},
publisher = {Association for Computing Machinery},
address = {New York, NY, USA},
volume = {28},
number = {4},
issn = {1084-4309},
journal = {ACM TODAES},
month = {may},
articleno = {52},
numpages = {31}
}

@INPROCEEDINGS{hls20,
  author={Kwon, Jihye and Carloni, Luca P.},
  booktitle={2020 ACM/IEEE 2nd Workshop on Machine Learning for CAD (MLCAD)}, 
  title={Transfer Learning for Design-Space Exploration with High-Level Synthesis}, 
  year={2020},
  volume={},
  number={},
  pages={163-168}}

@article{hls22,
author = {Wang, Zi and Schafer, Benjamin Carrion},
title = {Learning from the Past: Efficient High-level Synthesis Design Space Exploration for FPGAs},
year = {2022},
issue_date = {July 2022},
publisher = {Association for Computing Machinery},
address = {New York, NY, USA},
volume = {27},
number = {4},
issn = {1084-4309},

journal = {ACM Trans. Des. Autom. Electron. Syst.},
month = {feb},
articleno = {34},
numpages = {23}
}

@INPROCEEDINGS{ls19,
  author={Neto, Walter Lau and Austin, Max and Temple, Scott and Amaru, Luca and Tang, Xifan and Gaillardon, Pierre-Emmanuel},
  booktitle={2019 IEEE/ACM International Conference on Computer-Aided Design (ICCAD)}, 
  title={LSOracle: a Logic Synthesis Framework Driven by Artificial Intelligence: Invited Paper}, 
  year={2019},
  volume={},
  number={},
  pages={1-6}}

@INPROCEEDINGS{lo18,
  author={Haaswijk, Winston and Collins, Edo and Seguin, Benoit and Soeken, Mathias and Kaplan, Frédéric and Süsstrunk, Sabine and De Micheli, Giovanni},
  booktitle={2018 IEEE International Symposium on Circuits and Systems (ISCAS)}, 
  title={Deep Learning for Logic Optimization Algorithms}, 
  year={2018},
  volume={},
  number={},
  pages={1-4}}

@INPROCEEDINGS{gancts19,
  author={Lu, Yi-Chen and Lee, Jeehyun and Agnesina, Anthony and Samadi, Kambiz and Lim, Sung Kyu},
  booktitle={2019 IEEE/ACM International Conference on Computer-Aided Design (ICCAD)}, 
  title={GAN-CTS: A Generative Adversarial Framework for Clock Tree Prediction and Optimization}, 
  year={2019},
  volume={},
  number={},
  pages={1-8}}

@misc{gr19,
      title={A Deep Reinforcement Learning Approach for Global Routing}, 
      author={Haiguang Liao and Wentai Zhang and Xuliang Dong and Barnabas Poczos and Kenji Shimada and Levent Burak Kara},
      year={2019},
      eprint={1906.08809},
      archivePrefix={arXiv},
      primaryClass={cs.LG}
}

@misc{maskplace22,
      title={MaskPlace: Fast Chip Placement via Reinforced Visual Representation Learning}, 
      author={Yao Lai and Yao Mu and Ping Luo},
      year={2022},
      eprint={2211.13382},
      archivePrefix={arXiv},
      primaryClass={cs.CV}
}

@article{verif23,
title = {A test vector selection method based on machine learning for efficient presilicon verification},
journal = {Expert Systems with Applications},
volume = {224},
pages = {120056},
year = {2023},
issn = {0957-4174},
url = {https://www.sciencedirect.com/science/article/pii/S0957417423005584},
author = {Hyeong Gu Lim and Jaeyeon Jang and Byung Kook Ju and Jae Woo Ko and Chang Ouk Kim}
}

@unpublished{asml,
title= {Machine Learning in Computational Lithography},
author = {Yu Cao},
year = {2019},
URL= {https://www.ebeam.org/docs/SPIE2019-yu-cao.pdf}
}

@unpublished{opencores,
title= {OpenCores},
URL= {https://opencores.org}
}

@unpublished{iwls,
title= {IWLS 2005},
URL={https://iwls.org/iwls2005/benchmarks.html}
}

@unpublished{mcnc,
title= {MCNC},
URL={http://vlsicad.eecs.umich.edu/BK/MCNCbench}
}

@unpublished{gsrc,
title= {GSRC},
URL={http://vlsicad.eecs.umich.edu/BK/GSRCbench}
}

@unpublished{ispd02,
title= {ISPD-IBM},
URL={http://vlsicad.eecs.umich.edu/BK/ISPD02bench/#The_Benchmarks}
}

@unpublished{bookshelf,
title= {Bookshelf},
URL={http://vlsicad.eecs.umich.edu/BK/overview.htm}
}

@INPROCEEDINGS{adya01,
  author={Adya, S.N. and Markov, I.L.},
  booktitle={ICCD 2001}, 
  title={Fixed-outline floorplanning through better local search}, 
  year={2001},
  volume={},
  number={},
  pages={328-334}}

@misc{ct2021,
  title = {{Circuit Training}: An open-source framework for generating chip
  floor plans with distributed deep reinforcement learning.},
  author = {Guadarrama, Sergio and Yue, Summer and Boyd, Toby and Jiang, Joe
  Wenjie and Songhori, Ebrahim and Tam, Terence and Mirhoseini, Azalia},
  url = "https://github.com/google_research/circuit_training",
  year = 2021,
  note = "[Online; accessed 21-December-2021]"
}

@INPROCEEDINGS{eyecharts,
  author={Gupta, Puneet and Kahng, Andrew B. and Kasibhatla, Amarnath and Sharma, Puneet},
  booktitle={Design Automation Conference}, 
  title={Eyecharts: Constructive benchmarking of gate sizing heuristics}, 
  year={2010},
  volume={},
  number={},
  pages={597-602}}

@INPROCEEDINGS{gt14,
  author={Han, Seung-Soo and Kahng, Andrew B. and Nath, Siddhartha and Vydyanathan, Ashok S.},
  booktitle={DATE'14}, 
  title={A deep learning methodology to proliferate golden signoff timing}, 
  year={2014},
  volume={},
  number={},
  pages={1-6}}

@INPROCEEDINGS{cornerpred,
  author={Kahng, Andrew B. and Mallappa, Uday and Saul, Lawrence and Tong, Shangyuan},
  booktitle={DATE'19}, 
  title={"Unobserved Corner" Prediction: Reducing Timing Analysis Effort for Faster Design Convergence in Advanced-Node Design}, 
  year={2019},
  volume={},
  number={},
  pages={168-173}}

@ARTICLE{probe2,
  author={Cheng, Chung-Kuan and Kahng, Andrew B. and Kim, Hayoung and Kim, Minsoo and Lee, Daeyeal and Park, Dongwon and Woo, Mingyu},
  journal={IEEE Transactions on Computer-Aided Design of Integrated Circuits and Systems}, 
  title={PROBE2.0: A Systematic Framework for Routability Assessment From Technology to Design in Advanced Nodes}, 
  year={2022},
  volume={41},
  number={5},
  pages={1495-1508}}

@ARTICLE{kim23,
  author={Kim, Daeyeon and Lee, Sung-Yun and Min, Kyungjun and Kang, Seokhyeong},
  journal={IEEE Transactions on Computer-Aided Design of Integrated Circuits and Systems}, 
  title={Construction of Realistic Place-and-Route Benchmarks for Machine Learning Applications}, 
  year={2023},
  volume={42},
  number={6},
  pages={2030-2042}}

@article{fault21,
title = {Fault Detection based on Deep Learning for Digital VLSI Circuits},
journal = {Procedia Computer Science},
volume = {194},
pages = {122-131},
year = {2021},
note = {18th International Learning \& Technology Conference 2021},
issn = {1877-0509},
url = {https://www.sciencedirect.com/science/article/pii/S1877050921021062},
author = {Lamya Gaber and Aziza I. Hussein and Mohammed Moness},

}

@inproceedings{ispd12,
author = {Ozdal, Muhammet Mustafa and Amin, Chirayu and Ayupov, Andrey and Burns, Steven and Wilke, Gustavo and Zhuo, Cheng},
title = {The ISPD-2012 discrete cell sizing contest and benchmark suite},
year = {2012},
isbn = {9781450311670},
publisher = {Association for Computing Machinery},
address = {New York, NY, USA},
url = {https://doi.org/10.1145/2160916.2160950},

pages = {161–164},
numpages = {4},

series = {ISPD '12}
}

@inproceedings{ispd15,
author = {Bustany, Ismail S. and Chinnery, David and Shinnerl, Joseph R. and Yutsis, Vladimir},
title = {ISPD 2015 Benchmarks with Fence Regions and Routing Blockages for Detailed-Routing-Driven Placement},
year = {2015},
isbn = {9781450333993},
publisher = {Association for Computing Machinery},
pages = {157–164},
numpages = {8},

series = {ISPD '15}
}

@INPROCEEDINGS{iccad17,
  author={Darav, Nima Karimpour and Bustany, Ismail S. and Kennings, Andrew and Mamidi, Ravi},
  booktitle={2017 IEEE/ACM International Conference on Computer-Aided Design (ICCAD)}, 
  title={ICCAD-2017 CAD contest in multi-deck standard cell legalization and benchmarks}, 
  year={2017},
  volume={},
  number={},
  pages={867-871}}

@inproceedings{ispd18,
author = {Mantik, Stefanus and Posser, Gracieli and Chow, Wing-Kai and Ding, Yixiao and Liu, Wen-Hao},
title = {ISPD 2018 Initial Detailed Routing Contest and Benchmarks},
year = {2018},
isbn = {9781450356268},
publisher = {Association for Computing Machinery},
address = {New York, NY, USA},
booktitle = {Proceedings of the 2018 International Symposium on Physical Design},
pages = {140–143},
numpages = {4},
location = {Monterey, California, USA},
series = {ISPD '18}
}

@ARTICLE{adya03,
  author={Adya, S.N. and Markov, I.L.},
  journal={IEEE Transactions on Very Large Scale Integration (VLSI) Systems}, 
  title={Fixed-outline floorplanning: enabling hierarchical design}, 
  year={2003},
  volume={11},
  number={6},
  pages={1120-1135}}

@ARTICLE{btree06,
  author={Tung-Chieh Chen and Yao-Wen Chang},
  journal={IEEE Transactions on Computer-Aided Design of Integrated Circuits and Systems}, 
  title={Modern floorplanning based on B/sup */-tree and fast simulated annealing}, 
  year={2006},
  volume={25},
  number={4},
  pages={637-650}}

@article{gplanner22,
author = {Liu, Yiting and Ju, Ziyi and Li, Zhengming and Dong, Mingzhi and Zhou, Hai and Wang, Jia and Yang, Fan and Zeng, Xuan and Shang, Li},
title = {GraphPlanner: Floorplanning with Graph Neural Network},
year = {2022},
issue_date = {March 2023},
publisher = {Association for Computing Machinery},
address = {New York, NY, USA},
volume = {28},
number = {2},
journal = {ACM Trans. Des. Autom. Electron. Syst.},
month = {dec},
articleno = {21},
numpages = {24}
}

@inproceedings{kdd22,
author = {Amini, Mohammad and Zhang, Zhanguang and Penmetsa, Surya and Zhang, Yingxue and Hao, Jianye and Liu, Wulong},
title = {Generalizable Floorplanner through Corner Block List Representation and Hypergraph Embedding},
year = {2022},
isbn = {9781450393850},
publisher = {Association for Computing Machinery},
pages = {2692–2702},
series = {KDD '22}
}

@article{gfp22,
author = {Xu, Qi and Geng, Hao and Chen, Song and Yuan, Bo and Zhuo, Cheng and Kang, Yi and Wen, Xiaoqing},
title = {GoodFloorplan: Graph Convolutional Network and Reinforcement Learning-Based Floorplanning},
year = {2022},
issue_date = {Oct. 2022},
publisher = {IEEE Press},
volume = {41},
number = {10},
journal = {Trans. Comp.-Aided Des. Integ. Cir. Sys.},
month = {oct},
pages = {3492–3502},
numpages = {11}
}
\end{document}